\documentclass[aps,prb,reprint,superscriptaddress]{revtex4-1}

\usepackage{graphicx}
\usepackage[dvipsnames]{xcolor}

\begin{document}

\title{Band structure of CuMnAs probed by optical and
  photoemission spectroscopy}

\author{M. Veis}
\affiliation{Charles University, Faculty of Mathematics and Physics,
  Ke Karlovu 5, Praha 2, Czech Republic}

\author{J. Min\'ar}
\affiliation{New Technologies-Research Center University of West
  Bohemia, Plze{\v n}, Czech Republic}

\author{G. Steciuk}
\affiliation{Institute of Physics, Academy of Science of the Czech Republic, 
Cukrovarnick\'a 10, Praha 6, Czech Republic}

\author{L. Palatinus}
\affiliation{Institute of Physics, Academy of Science of the Czech Republic, 
Cukrovarnick\'a 10, Praha 6, Czech Republic}

\author{C. Rinaldi}
\affiliation{Department of Physics, Politecnico di Milano,
  via G. Colombo 81, 20133, Milano, Italy}

\author{M. Cantoni}
\affiliation{Department of Physics, Politecnico di Milano,
  via G. Colombo 81, 20133, Milano, Italy}

\author{D. Kriegner}
\affiliation{Institute of Physics, Academy of Science of the Czech Republic, 
Cukrovarnick\'a 10, Praha 6, Czech Republic}
\affiliation{Charles University, Faculty of Mathematics and Physics,
  Ke Karlovu 5, Praha 2, Czech Republic}

\author{K.K.~Tikui\v sis}
\affiliation{Charles University, Faculty of Mathematics and Physics,
  Ke Karlovu 5, Praha 2, Czech Republic}

\author{J. Hamrle}
\affiliation{Charles University, Faculty of Mathematics and Physics,
  Ke Karlovu 5, Praha 2, Czech Republic}

\author{M. Zahradn\'\i{}k}
\affiliation{Charles University, Faculty of Mathematics and Physics,
  Ke Karlovu 5, Praha 2, Czech Republic}

\author{R. Anto\v s}
\affiliation{Charles University, Faculty of Mathematics and Physics,
  Ke Karlovu 5, Praha 2, Czech Republic}

\author{J. \v Zelezn\'y}
\affiliation{Institute of Physics, Academy of Science of the Czech Republic, 
Cukrovarnick\'a 10, Praha 6, Czech Republic}

\author{L. \v Smejkal}
\affiliation{Institute of Physics, Academy of Science of the Czech Republic, 
Cukrovarnick\'a 10, Praha 6, Czech Republic}

\author{P.~Wadley}
\affiliation{School of Physics and Astronomy, University of
  Nottingham, Nottingham NG7 2RD, United Kingdom}

\author{R.P.~Campion}
\affiliation{School of Physics and Astronomy, University of
  Nottingham, Nottingham NG7 2RD, United Kingdom}

\author{\hbox{C. Frontera}}
\affiliation{Institut de Ci\`encia de Materials de Barcelona
  (ICMAB-CSIC), Campus Universitari de Bellaterra, Cerdanyola del Vall\`es,
  08193 Spain}

\author{K.~Uhl\'\i\v rov\'a}
\affiliation{Charles University, Faculty of Mathematics and Physics,
  Ke Karlovu 5, Praha 2, Czech Republic}

\author{T. Ducho\v n}
\affiliation{Charles University, Faculty of Mathematics and Physics,
  Department of Surface and Plasma Science, V Hole\v sovi\v ck\' ach 2, 18000
  Praha 8, Czech Republic}

\author{P. Ku\v zel}
\affiliation{Institute of Physics, Academy of Science of the Czech Republic, 
Na Slovance 1999/2, Praha 8, Czech Republic}

\author{V. Nov\'ak}
\affiliation{Institute of Physics, Academy of Science of the Czech Republic, 
Cukrovarnick\'a 10, Praha 6, Czech Republic}

\author{T. Jungwirth}
\affiliation{Institute of Physics, Academy of Science of the Czech Republic, 
Cukrovarnick\'a 10, Praha 6, Czech Republic}
\affiliation{School of Physics and Astronomy, University of
  Nottingham, Nottingham NG7 2RD, United Kingdom}

\author{K. V\'yborn\'y}
\affiliation{Institute of Physics, Academy of Science of the Czech Republic, 
Cukrovarnick\'a 10, Praha 6, Czech Republic}
  
\date{Dec01, 2017 version + SI}

\begin{abstract}
Tetragonal phase of CuMnAs progressively appears as one of the key materials
for antiferromagnetic spintronics due to efficient current-induced spin-orbit
torques whose existence can be directly inferred from crystal symmetry.
Theoretical understanding of spintronic phenomena in this material,
however, relies on the detailed knowledge of electronic structure
(band structure and corresponding wave functions)
which has so far been tested only to a limited extent. We show that AC
permittivity (obtained from ellipsometry) and UV photoelectron spectra
agree with density functional calculations. Together with the x-ray
diffraction and precession electron diffraction tomography, our
analysis confirms recent theoretical claim [Phys.Rev.B 96, 094406 (2017)]
that copper atoms occupy lattice positions in the basal plane of
the tetragonal unit cell. 
\end{abstract}

\pacs{later}

\maketitle

%\section{Introduction}

Magnetic moments in antiferromagnets have been notoriously difficult to
manipulate. With the exception of materials having low
N\'eel temperature and small magnetic anisotropy, very strong magnetic
fields must be applied. Such fields would be too strong to be of any
practical use and, moreover, they can never be applied as locally as
electric pulses. Recently, an alternative manipulation mechanism has been
proposed\cite{Zelezny:2014_a} which relies on current-induced spin-orbit
torques (SOTs) acting in the bulk of the antiferromagnetic material.
They result from a build-up of staggered spin polarisation (i.e. the one
which alternates sign on two magnetic sublattices) in response to an
applied uniform electric current; such polarisation can be
calculated in the framework of linear response to electric
field.\cite{Edelstein:1989_a,Bernevig:2005_a}
A prediction of sizable SOT in CuMnAs has soon been experimentally
confirmed\cite{Wadley:2016_a} and prototype memories where the writing is
done using SOT have been demonstrated.\cite{Olejnik:2017_a} Devices
based on thin films of CuMnAs thus claim a prominent role within
the fast developing field of antiferromagnetic
spintronics.\cite{Baltz:2017_a,Jungwirth:2016_a}

Quantitative modelling of SOT (and many other material-specific quantities)
relies on a detailed knowledge of the
electronic structure.\cite{Li:2015_a} While well-established ab initio
methods have been used for this purpose, little effort has so far been
dedicated to validating the band structure in terms of comparing
calculated and measured spectral properties.\cite{note2} We fill this
gap by exploring the complex AC permittivity in the optical range and
photoemission spectroscopy in the UV range (UPS) and comparing them to
density functional theory (DFT) calculations. We find a good agreement
between the experimental data and the calculated properties provided
the electronic correlations are treated beyond DFT, using Hubbard model
characterised by an on-site repulsion $U$ on Mn $3d$ orbitals.
Moreover, we demonstrate that
the AC permittivity in the optical range can be used to discern different
phases of CuMnAs. Focusing on the tetragonal phase of
CuMnAs,\cite{Wadley:2013_b} we corroborate analysis of our spectral
measurements by precession electron diffraction tomography (PEDT),
which points to a phase recently claimed to have the lowest
theoretically calculated total energy.\cite{Maca:2017_a} 

%\section{Samples} 

The studied thin films of CuMnAs were prepared by molecular
beam epitaxy (MBE; we followed procedures described in
Ref.~\onlinecite{Wadley:2013_b}). We performed the standard
x-ray diffraction (XRD) structural characterisation and PEDT.
Ellipsometry was carried out on a nominally 20~nm thick layer, while
PEDT was applied to 150~nm thick layers, both grown on a GaP(001) substrate.
Photoemission spectra in the UV range (UPS) were obtained for a
130 nm thick sample grown on a GaAs(001) substrate.

\begin{figure}
\includegraphics[scale=0.45]{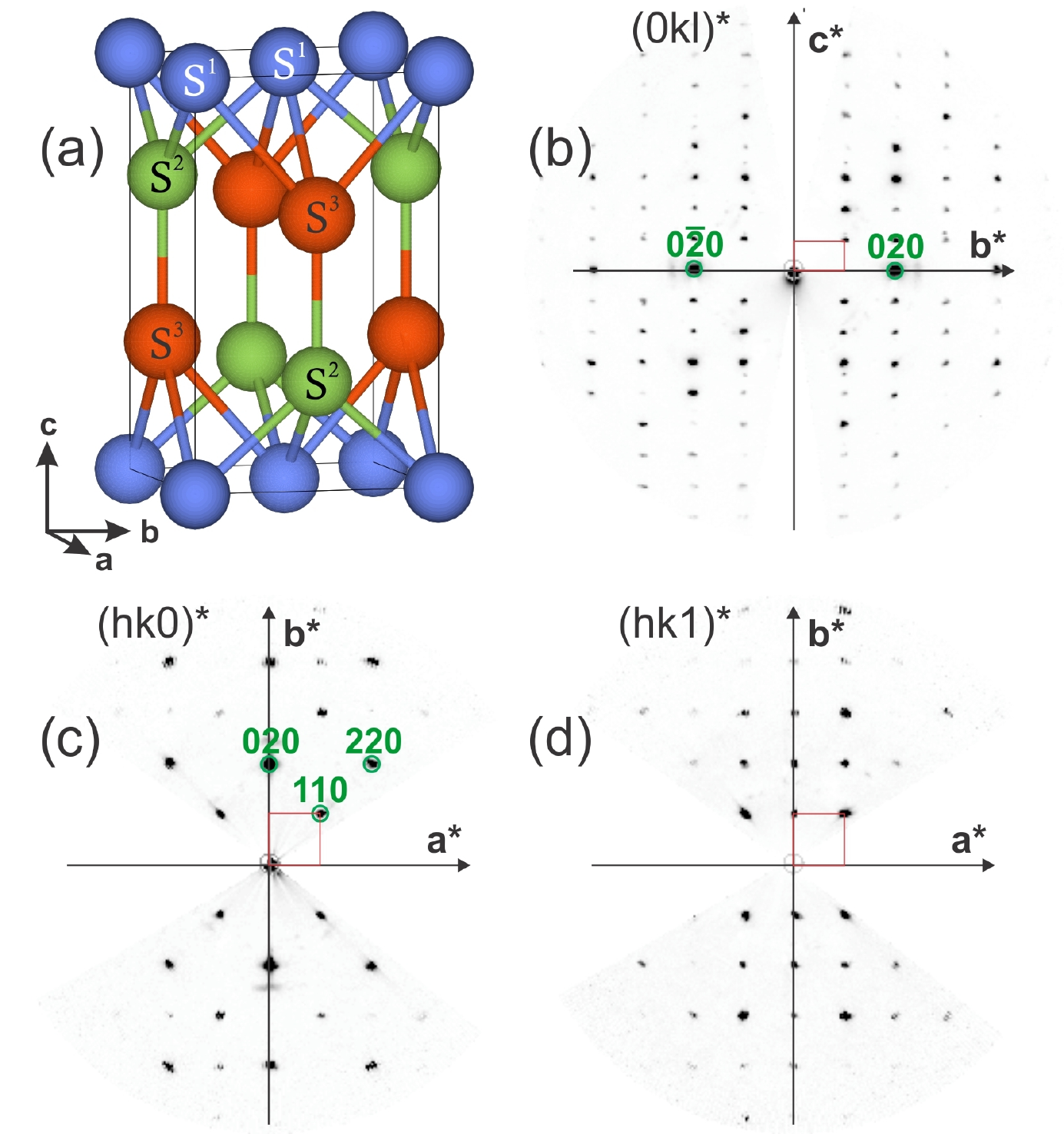}
\caption{(a) Structure of the tetragonal phase of CuMnAs. In RTP, Cu/As/Mn
  atoms occupy the sites $S^1/S^2/S^3$. In the inverted phase, Cu
  and Mn atoms are swapped. (b-d) Electron diffraction
  (PEDT) patterns: (0kl)*, (hk0)* and (hk1)* sections of the reciprocal space
  reconstructed from  PEDT data (PETS program\cite{Palatinus:2011_a}).
  The conditions $h+k=2n$ on $hk0$, $h=2n$ on $h00$ and $k=2n$ on
  $0k0$ are characteristic of the n~glide.}
\label{fig-01}
\end{figure}

Both XRD and PEDT confirm the tetragonal crystal structure shown in
Fig.~\ref{fig-01} with space group P4/nmm.
Regarding the occupancy of lattice sites labelled $S^1$, $S^2$ and
$S^3$ in the Figure, neither x-ray nor electron 
diffraction are very efficient in distinguishing manganese and copper
atoms because of their similar scattering powers; the latter method,
however, does provide some advantage over the former one as we show
below. In this work, we consider theoretically
two tetragonal phases which are defined as follows: the first
structure has copper atoms at the basal positions $S^1$ (Wyckoff
position $2a$) and As/Mn at $S^2/S^3$ (Wyckoff position $2c$). We will
refer to this as the reference tetragonal phase (RTP). The second,
inverted structure is obtained by swapping Mn and Cu so that the basal
positions $S^1$ are occupied by manganese. Regarding magnetic structure
of the latter phase, we only consider the case of antiferromagnetic ordering
within the basal plane where the unit cell contains six atoms.

Using XRD, we find lattice constants differing by less than 1\% for
samples grown on GaP and GaAs (e.g. $a=0.3853$~nm and $a=0.3820$~nm at
room temperature, respectively). For the purpose of effects considered
in this work, such differences lead to negligible changes in observed
spectra, which renders, within the scope of this article, all our thin film
samples interchangeable. Further details about x-ray characterisation
can be found in the Supplementary information\cite{SI} (Sec.~I) and we
now turn our attention to the electron diffraction analysis.

%\section{Electron diffraction analysis}

For PEDT characterisation, a cross section of the CuMnAs/GaP(001)
thin film was prepared by mechanic polishing followed by ion
milling. Four PEDT data sets were recorded on several
parts of the film using a Philips CM120 electron transmission
microscope ($V_{acc}=120$~kV, LaB$_6$) with the precession device
Nanomegas Digistar and a side-mounted CCD camera Olympus Veleta with
14bit dynamic range. The precession angle and the tilt step of the
goniometer were both set to 1~degree.
The data were analysed using the computer programs
PETS\cite{Palatinus:2011_a} and JANA2006.\cite{Petricek:2014_a}

Tetragonal structure as shown in Fig.~\ref{fig-01}(a) was
confirmed by the PEDT data. Extinction conditions observed on
the sections of the reciprocal space shown in Figs.~\ref{fig-01}(b-d)
are compatible with the space group $P$4/nmm. RTP was used as a starting
structure and refined from the PEDT data using the dynamical theory of
diffraction ("dynamical refinement") according to
\hbox{Refs.~\onlinecite{Palatinus:2015_a,Palatinus:2015_b}}.
All four PEDT data sets were combined in order to increase the statistics
of the refinement and the coverage of the reciprocal space. Results of the
refinement are summarised in Table~\ref{tab-01}. We measured almost
five thousand reflections in all data sets (N$_{all}$) and found
N$_{obs}$ reflections with a significant intensity. Among model
parameters, there are seven structural parameters (two $z/c$ parameters,
three displacement parameters and two occupancy factors), an average
thickness of the analysed area for each of the four datasets and one
scaling parameter per each experimental diffraction pattern giving in
total N$_{param}\ll \mbox{N}_{obs}$ optimised parameters. Note that
data-to-parameter ratio N$_{all}$/N$_{param}>10$ is required for a
reliable structure determination. The quality of the fit is demonstrated
by the $R$-value\cite{note1} of 10.46 and only slightly larger
weighted $R$-value. For atomic positions $z/c$, we obtain values in a good
agreement with the corresponding values inferred from x-ray analysis.\cite{SI}
The occupancy of $S^3$ is found significantly different from one
suggesting that our samples are copper-rich.

\begin{table*}
  \caption{Crystallographic and dynamical refinement parameters
    of the RTP and the inverted tetragonal phases, both of
    $P$4/nmm (No. 129) space group.    For RTP: Cu ($S^1$)
    occupies the Wyckoff position 2a ($\frac{1}{2}$,$\frac{1}{2}$,0),
    Mn ($S^3$) and As ($S^2$) occupy positions 2c
          (0,$\frac{1}{2}$,$ z$). In the inverted
          tetragonal phase $S^1$=Mn and $S^3$=Cu.}
        \label{tab-01}  %{PEDT_dyn}
        \renewcommand{\arraystretch}{0.7}
{\footnotesize
        \begin{center}
                        \begin{tabular}{lcccccccc}
                                \hline
                                 \multicolumn{9}{l}{Structural parameters:}  \\ 
                                 &\multicolumn{2}{c}{$z/c$} & \multicolumn{3}{c}
{occupancies} & \multicolumn{3}{c}{ADPs (iso.)[\AA$^2$]} \\ 
                                &$S^2$ &   $S^3$ & $S^1$ & $S^2$ & $S^3$ & $S^1$
 & $S^2$ & $S^3$ \\ 
                                RTP: &0.2627(2) &  0.6628(2) &\hskip1cm 0.995(8)$\sim$1 
&1  &0.869(7) & 0.0147(4) & 0.0121(3)& 0.0123(4)  \\ 
                                inverted: &0.2627(2)& 0.6624(2) & \hskip.7cm 0.870(6) &1 &0.949(8) & 0.0052(4) & 0.0124(3)& 0.0205(5)  \\ 
                                 \multicolumn{9}{l}{Refinement parameters:}  \\ 
                                RTP:  &\multicolumn{2}{c}{N$_{param.}$=338;}  &\multicolumn{2}{c}{N$_{obs/all}$.=3768/4189;}  &\multicolumn{2}{c}{R$_{obs}$=10.15;} & \multicolumn{2}{c}{wR$_{all}$=11.86} \\
                                inverted: &\multicolumn{2}{c}{N$_{param.}$=338;}  &\multicolumn{2}{c}{N$_{obs/all}$.=3767/4190;}  &\multicolumn{2}{c}{R$_{obs}$=10.68;} & \multicolumn{2}{c}{wR$_{all}$=12.51} \\ \hline
                        
                        \end{tabular}
        \end{center}}
        \normalsize
        
\end{table*}

The key added value of PEDT in the context of this study is the
ability to better distinguish RTP from the inverted structure, and to this end,
the isotropic displacement parameters U$_{iso}(S^1)$ and
U$_{iso}(S^3)$ (also known as ADP) are the most sensitive indicators.
If atomic types are correctly assigned to individual atomic positions,
their values should be approximately equal.
For the $S^1$ and $S^3$ sites in Fig.~\ref{fig-01}, the ADPs in the
RTP model % are close within error bars  ($0.001\mbox{ \AA}^2$) and
do have similar values, consistent with previous
studies.\cite{Wadley:2013_a,Nateprov:2011_a}  However,
they change unfavourably for the inverted tetragonal phase: the ADP
drops (increases) by about 65\% for the $S^1$ ($S^3$) site, respectively.
This result is consistent with
the higher electron atomic scattering amplitude of Cu ($f^{B}_{Cu}$)
relative to Mn ($f^{B}_{Mn}$). In other words, RTP seems more
consistent with the electron diffraction data. Note that also the
$R$-value in Tab.~\ref{tab-01} for the inverted structure is
appreciably larger than for RTP. 

%\section{Ellipsometry}

\begin{figure}
\includegraphics[scale=0.3]{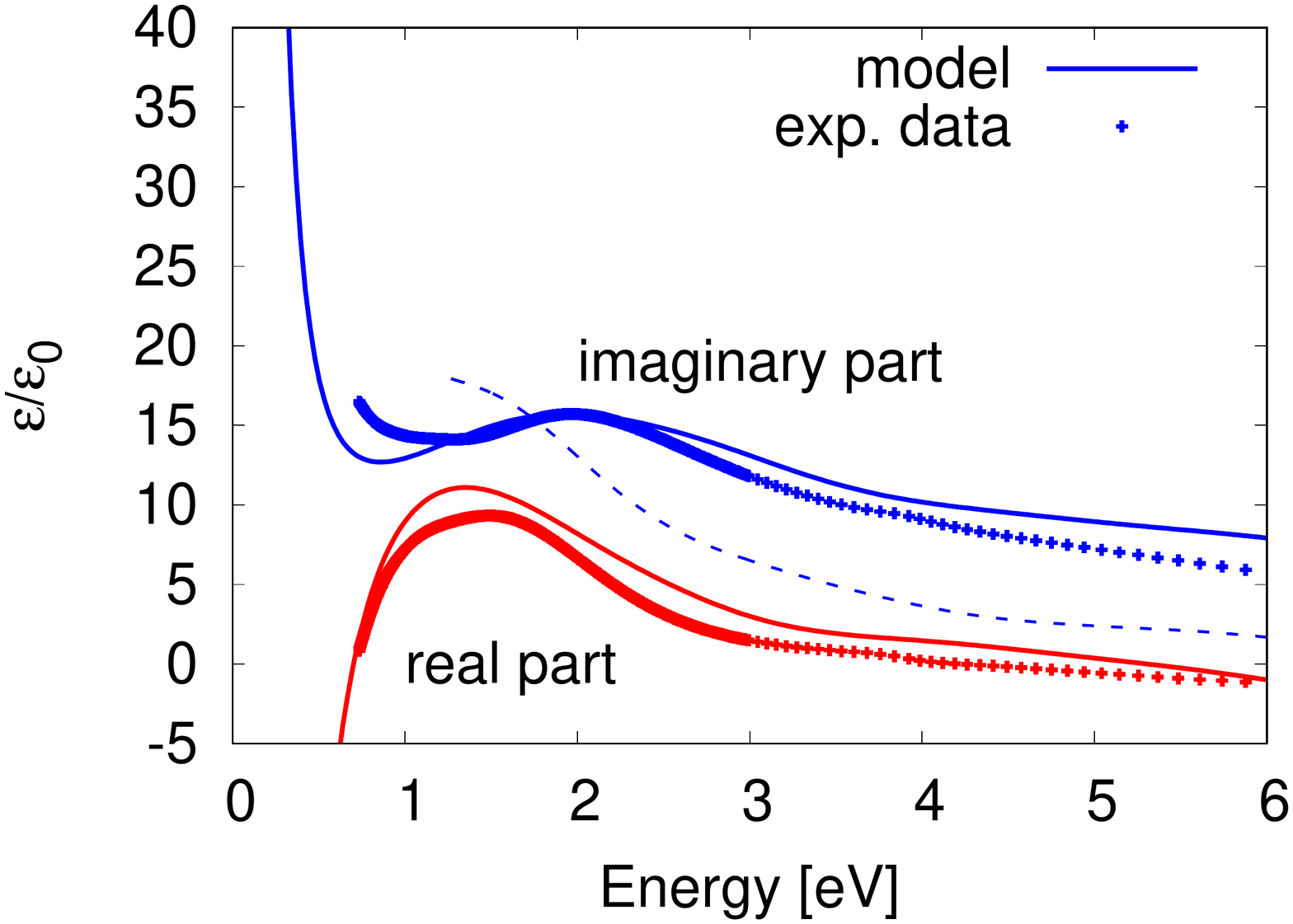}
\caption{AC permittivity of CuMnAs thin film determined by
  ellipsometry (for comparison, Im~$\epsilon/\epsilon_0$ for
  orthorhombic bulk CuMnAs is shown by the dotted line; see Supplementary
  information,\cite{SI} Sec.~III).
  The GGA+U model uses $U=1.7$~eV and $\Gamma=0.7$~eV for the
  interband part and $\hbar\omega_p=3.26$~eV and $\hbar/2\tau=120$~meV for the
  intraband contribution.}
\label{fig-02}
\end{figure}

A Mueller matrix ellipsometer JA Woollam RC2 was employed to acquire
experimental spectra of ellipsometric parameters $\Psi$ and
$\Delta$. To ensure a sufficiently large ensemble of
experimental data necessary for fitting, spectra were measured at
several angles of incidence ($55^\circ, 60^\circ, 65^\circ, 70^\circ$).
The experimental data were fitted using the Woollam CompleteEase software
starting with a model structure of nominally 20 nm thick \hbox{CuMnAs}
layer on GaP substrate and a surface oxide layer was accounted for,
which naturally occurs when the sample is exposed to air (see
Sec.~II of Supplementary information\cite{SI} for details).
Optical constants of GaP were taken from
literature,\cite{Aspnes:1983_a} while the permittivity of
\hbox{CuMnAs} was parametrised by a combination of Drude, Tauc-Lorentz and
three Lorentz functions. All parameters were fitted together with the
layer thickness ($l_{\mathrm{CuMnAs}}$) and surface roughness. The resulting
$l_{\mathrm{CuMnAs}}=22.6$~nm along with a negligible surface roughness
confirm the high level of sample growth control. Also, the mean square error
(MSE) was lower than~1, implying a rather robust fit whose result is shown in 
Fig.~\ref{fig-02} as experimental data.

Our DFT+U calculations\cite{Blaha:2001} for tetragonal CuMnAs
($a=0.3853$~nm, $c=0.6276$~nm) based on generalised gradient
approximation (GGA) with scalar-relativistic correction come quite
close to the experimental data (Fig.~\ref{fig-02}),
provided relatively large interband
broadening ($\Gamma=0.7$~eV) is used.  Such value is not
unprecedented\cite{Berlijn:2011_a} although still significantly larger
than $\hbar/2\tau$ implied by Drude-formula relaxation time
$\tau$ obtained from measured DC conductivity.
This said, one should be reminded that the intra- and interband
relaxation times are not required to be the same so that
parameters used for the model in Fig.~\ref{fig-02} are still
plausible. To estimate $\tau$, we used (apart from the
experimental resistivity\cite{Wadley:2013_b}) the ab-initio
calculated plasma frequency $\omega_p$. The model data plotted
in Fig.~\ref{fig-02} include also the intraband contribution (Drude
peak). From this point on, we will only be discussing the imaginary
part of permittivity since the Kramers-Kronig-related real part bears no
additional information.

\begin{figure}
\includegraphics[scale=0.2]{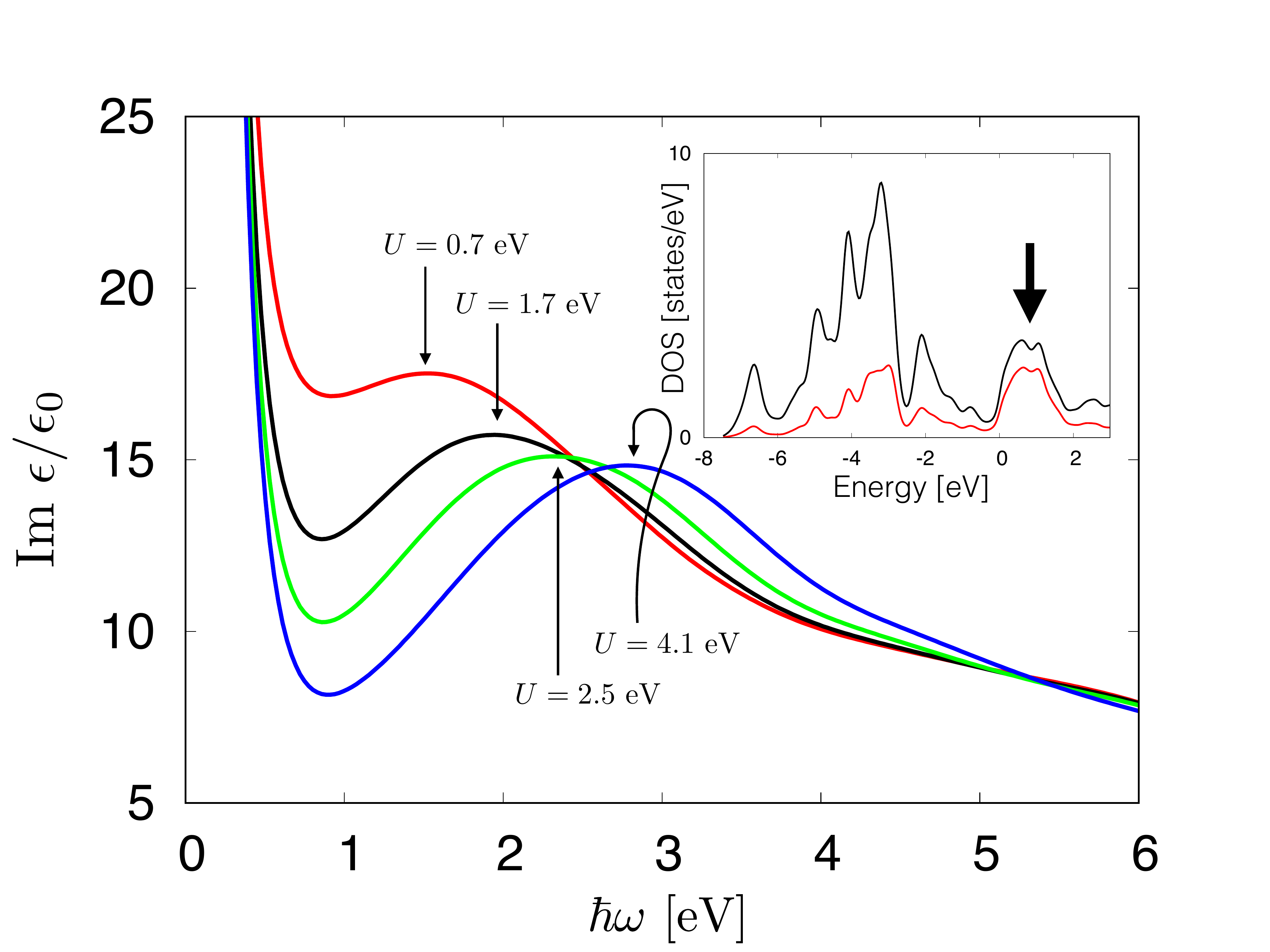} 
\caption{Imaginary part of relative permittivity calculated for several values
  of $U$. The peak shifts blue with increasing $U$,
  the arrows indicate the position of the maximum. {\em Inset:}
  density of states per spin (and Mn-partial density of states shown in red)
  with a Mn-dominated peak just above the Fermi level (taken as $E=0$).}
\label{fig-03}
\end{figure}

\begin{figure}
\includegraphics[scale=0.27]{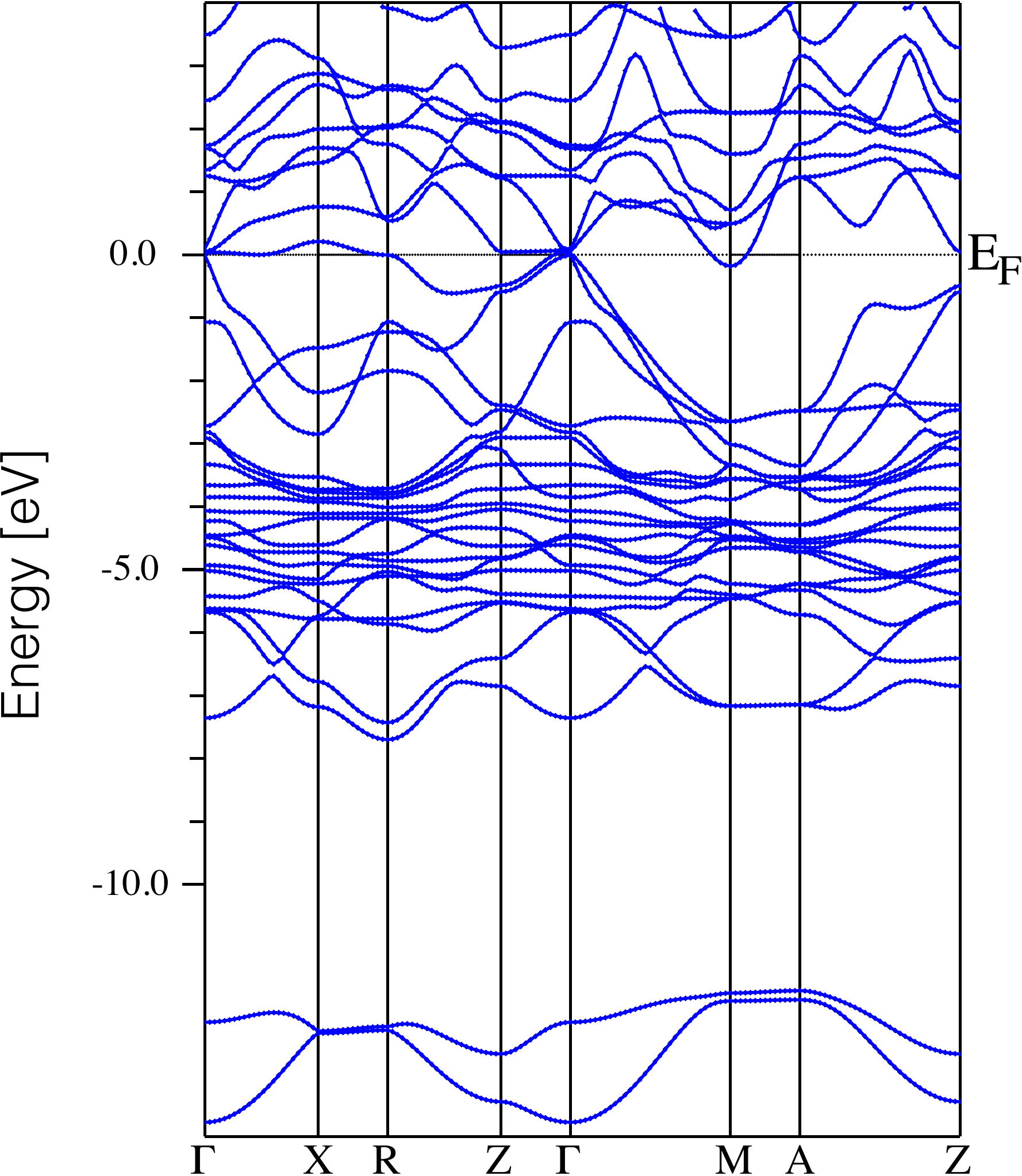}
\caption{Based on comparison between ellipsometry and GGA+U
  calculations,\cite{Blaha:2001}
  this band structure ($U=1.7$~eV) seems to describe
  well CuMnAs in tetragonal phase.}
\label{fig-04}
\end{figure}

Accounting for electron correlations turns out to be essential. We
use GGA+U with double-counting corrections to the DFT part treated in
the fully localised limit\cite{Anisimov:1993_a} (FLL) and find
the peak in imaginary part of the permittivity blue-shifting with
increasing value of $U$ (see Fig.~\ref{fig-03}). Its
experimentally determined position ($\hbar\omega\approx 2$~eV) is
recovered for $U=1.7$~eV and on the theoretical side, the peak stems from
unoccupied Mn states (indicated by an arrow in the inset of Fig.~\ref{fig-03}).
In Fig.~\ref{fig-04}, the corresponding band structure is shown. We
now also briefly discuss the effect of the parameter\cite{Anisimov:1993_a}
$J>0$. It causes the peak in Im~$\epsilon/\epsilon_0$
to shift to lower energies (in agreement with
replacing $U$ and $J$ by $U_{\mathrm{eff}}=U-J$ and $J=0$), and also it adds
some additional structure to the peak. The large interband broadening,
however, renders such effects unobservable. Based on ellipsometry
data, values of $U-J\approx 2$~eV therefore seem to give the best
results.

%\section{Angle--integrated photoemission}

\begin{figure}
\includegraphics[scale=0.30]{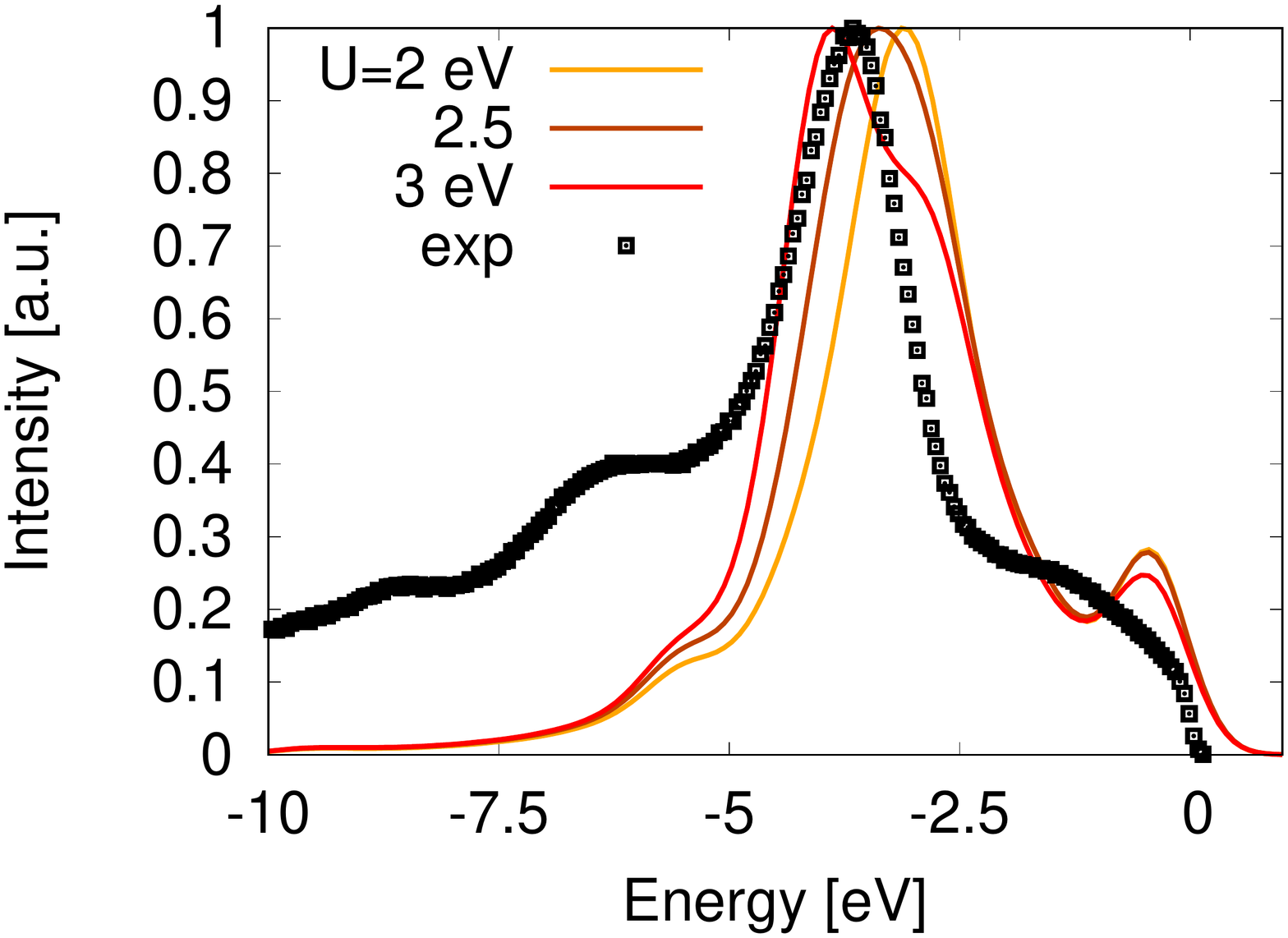}
\setlength{\unitlength}{1cm}
\put(-3.2,5){(A)}
\put(-2.6,2.7){(B)}
\put(-6,3){(C)}
\caption{Experimental angle-integrated photoemission (UPS, black squares)
  compared to the corresponding one-step model of photoemission.
  Theoretical data are shown for three values of 
  $U$, labelled features are described in the text.}
\label{fig-05}
\end{figure}

Photoemission spectra and also inverse photoemission
spectra (IPES) were measured for CuMnAs thin layers covered
originally (after growth) by an arsenic cap. This protective
layer was removed by Ar ion milling in the UHV environment for
UPS and IPES.\cite{Bertacco:2005_a} 
The cleanness of the surface was checked in-situ by X-ray
Photoemission Spectroscopy (XPS): the disappearance of core-level
peaks O~1s and C~1s indicates that the surface is clean (the residual
contamination is well below 1\% of surface coverage). The UPS spectrum
was recorded using a He lamp as excitation source (HeI-$\alpha=21.2$~eV)
and a hemispherical energy analyser Phoibos 150 (SPECS$^{\mathrm{TM}}$),
yielding an acceptance angle of $\sim 6^\circ$ and a field view of 1.4~mm$^2$.
The UPS spectrum of CuMnAs
is shown in Fig.~\ref{fig-05} by black squares. Results of the investigation
of empty states above the Fermi level by IPES is shown only in
the Supplementary Information\cite{SI} (Section~IV).
As a matter of fact, the calculated DOS above the Fermi level is less
sensitive to variations of $U$ and, moreover, fine details cannot be
accessed by IPES because of the large experimental
broadening\cite{Cantoni:1999_a} characteristic of these spectra. 

Photoemission spectroscopies access  the electronic
structure associated with top $\approx 1$~nm of the thin layer.\cite{Huf03}
In the simplest approximation, the measured angle-integrated UPS and IPES
should reflect rather directly the DOS. This approximation works
reasonably well in the high energy regime
(XPS) and led\cite{Zelezny:2016} to a larger estimate of $U$ around
4.5~eV. However, this approach ignores the influence of specific
matrix elements that, in general, introduce an energy- and
element-dependent weight to DOS. Also, in the
regime of low photon energies (as measured here), additional aspects
may have a very pronounced impact on the angle-integrated
photoemission spectra, for example final states as well as surface effects.

Here we used the recently developed full spin-density matrix formulation
for the photocurrent\cite{Pen76,Braun:2014_a} (see details in
Supplementary information\cite{SI}) within the relativistic
Korringa--Kohn--Rostoker Green function method. This method is
implemented in the SPR-KKR program package.\cite{Ebert:2011_a}
Regarding the value of $U$, we arrive at a somewhat different conclusion
than what was made in Ref.~\onlinecite{Zelezny:2016}. Nevertheless, the DOS
shown in the inset of Fig.~\ref{fig-03} still
provides a good means for interpreting, on an elementary level, both
the calculated and the measured spectra. They are dominated by the Mn
states located at $\approx 1$~eV and $-4$~eV (with respect to the
Fermi level), the latter having a significant admixture of Cu
states. The peak at $\approx -2$~eV with dominantly Cu character
is not visible in the UPS spectra, being probably hidden in the main peak of
the measured data. The three main features in experimental spectra are
labelled by capital letters in Fig.~\ref{fig-05}. It turns out that the main
strong peak (A) serves as the best test for calculated spectra and
their dependence on the value of $U$. As this value increases, the
peak blue-shifts and a match with the experiment occurs between
$2.5$~eV and 3~eV (see Fig.~\ref{fig-05}; note, however, that the
shoulder which develops in model calculations for $U=3$~eV is absent
in experimental data suggesting the plausibility of lower values
of $U$). Both this feature and (C)
which is also clearly visible in the model calculations, can be back
tracked to the Mn $d$-states which are shifted to higher binding
energies when $U$ increases. The broad peak (B) located close to
the Fermi level shows a strong surface character. We confirmed this
theoretically by modifying the surface barrier (see Sec.~IV. in Supplementary
information\cite{SI}). Given that the surface was probably damaged by
ion milling used to remove the cap, only little information about the
bulk electronic structure can be extracted from this part of UPS. The
level of agreement between UPS data and DFT+U calculations suggests
that aforementioned $U-J\approx 2$~eV is an acceptable value for
the band structure calculations (using the particular method
of Ref.~\onlinecite{Blaha:2001}).

%\section{Conclusions}

In conclusion, we presented optical spectra of the complex permittivity and
photoemission spectra in the UV regime of MBE-grown thin layers of CuMnAs,
which crystallise in the tetragonal structure, and demonstrated a good level
of agreement with the DFT+U calculations. Together with the dynamically refined
precession electron diffraction tomography, this agreement strongly
suggests that copper occupies the basal positions of the structure
($S^1$ in Fig.~\ref{fig-01}) and this confirms the conclusions of the
recent theoretical study of M\'aca et al.~[\onlinecite{Maca:2017_a}].

%\section{Acknowledgements}
We acknowledge discussions with R. Bertacco and support
from National Grid Infrastructure MetaCentrum
provided under the programme "Projects of Large Research, Development,
and Innovations Infrastructures" (CESNET LM2015042);
Grant Agency of the Czech Republic under No. 15-13436S;
CEDAMNF (CZ.02.1.01/0.0/0.0/15\_003/0000358) of the Czech ministry of
education (M\v SMT);
Cariplo Foundation, grant No. 2013-0726 (MAGISTER);
Spanish MINECO under MAT2015-67593-P project and the 'Severo Ochoa'
Programme (SEV-2015-0496);  EU FET Open RIA Grant no. 766566.

\newpage

\begin{widetext}

\section*{Supplementary information $\equiv$ ref. 12}

%%% fte-list  SI
% fig-01, fig-02, fig-03, fig-04, fig-05, fig-06
% tab-01
%

\section{x-ray characterisation}

X-ray scattering and diffraction measurements were used to check the
thin-film thickness and the tetragonal structure in Fig.~1 of the main text.
%Fig.~\ref{fig-04}.
The obtained parameters, described in detail below,
are summarised in Tab.~\ref{SI-tab-01} here, and can be compared to values
from Tab.~I in the main text. We first describe the X-ray
reflectivity analysis which yields the film thickness and roughness of
the sample which we compare to the analysis of ellipsometric
measurements. After that we turn
to the structural characterization complementary to the PEDT data
presented in the main text.

%\begin{figure}[b]
%\includegraphics[width=0.7\linewidth]{figs/p04.pdf}
%%\includegraphics[width=0.7\linewidth]{p04.pdf}
%\caption{The structure of tetragonal CuMnAs. Atomic site $S^2$ is
%  always occupied by As, sites $S^1$ and $S^3$ are occupied by Cu and Mn.}
%\label{fig-04}
%\end{figure}

\subsection{X-ray reflectivity analysis of the film thickness}

X-ray reflectometry (XRR) was recorded with a Rigaku Smartlab rotating
anode system using monochromatic X-ray photons with CuK$_{\alpha1}$
wavelength. Since the refractive index of materials is below unity
total external reflection occurs under small incidence angles. Above
the critical angle where the X-rays penetrate the film material, Kiessig
fringes due to the interference of X-rays reflected at the top and
bottom interface of the thin film correspond to the film thickness and
can be modeled by the Parrat formalism.\cite{Holy:2004} Experimental
data in comparison with simulations are shown in Fig.~\ref{SI-fig-02}.
The simulations yield a total film thickness of 22.7~nm and a surface
and interface roughness of around 0.4~nm. We note that the X-ray
reflectivity analysis was performed after the sample was exposed to
air for longer time and therefore some oxide, present on the surface
as discussed below, was included in the XRR simulations. For
comparison with values determined by ellipsometry (see main text and
the discussion in Sec.~II below) we therefore use the total
thickness. Given that different pieces of wafer were used for
individual measurements (i.e. the same growth run but different
storage conditions), we find an excellent agreement.

\begin{figure}[b]
\includegraphics[width=0.7\linewidth]{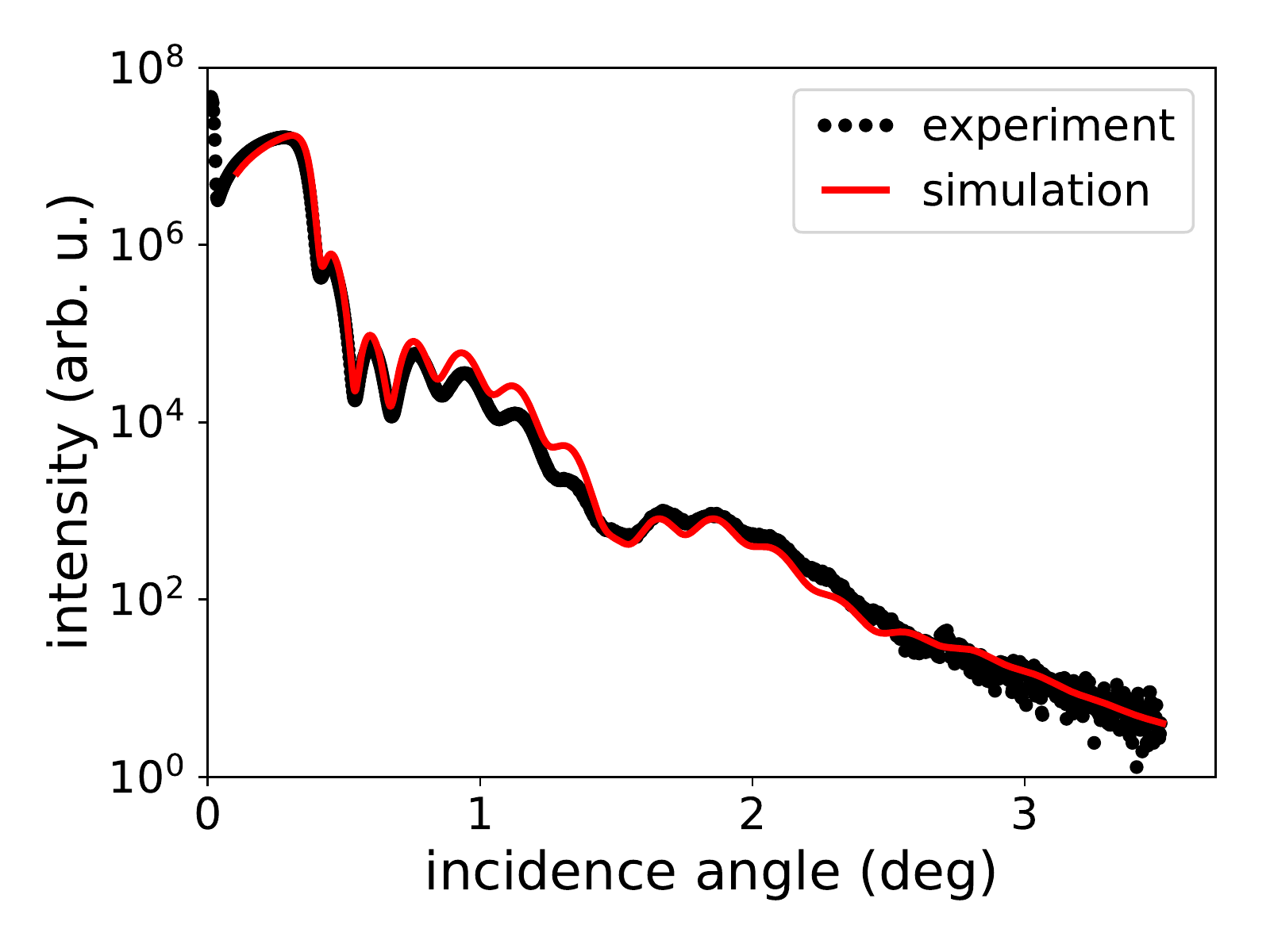}
\caption{X-ray reflectivity data and simulations of the sample investigated by
ellipsometry.}
\label{SI-fig-02}
\end{figure}

\subsection{X-ray diffraction analysis of the tetragonal structure}

Due to large penetration depth of x-ray photons, X-ray diffraction (XRD) probes
the full film depth including the substrate below --- the advantage being
that no sample preparation is required for XRD measurements, which therefore
naturally complement the PEDT data in the main text. Despite
certain other advantages, PEDT requires a complex
specimen preparation. Further, XRD has a higher precision for lattice parameter
measurements; thin films grown on GaP(001) yield 
lattice parameters specified in Tab.~\ref{SI-tab-01}. These measurements were
performed using the aforementioned Rigaku Smartlab rotating anode system using
monochromatic X-ray photons with CuK$_{\alpha1}$ and a linear detector.

The determination of the atomic structure of the thin films was performed as
described in detail in Ref.~\onlinecite{SI-Wadley:2013_a}. For that purpose we
used a Bruker D8-Discover diffractometer equipped with a V{\aa}ntec-500 area
detector. The difference between lattice parameter of CuMnAs ($\sim 6.3$~\AA)
and GaP ($\sim 5.45$~\AA) warrants that no significant overlapping between
diffraction peaks from substrate and from film takes place.  The total number
of diffraction peaks measured was 44. The number of observable diffractions is
therefore considerable lower as in the PEDT data in the main text. This is
compensated by easier modeling which in total only requires 5 structural
parameters as opposed to observables (27 independent reflections) and
a single one overall factor.
%%% Here, entering in atomic occupancies would require talking about EPMA, 
%%% as it is the starting point for refinements
The structural refinement includes an overall temperature factor, the two $z/c$
positions of $S_2$ and $S_3$, reported in the Tab.~\ref{SI-tab-01}. The shorter
range in $Q$-space available for the long wavelength of Cu-K$\alpha$ (in
comparison with electrons), hinders the refinement of individual temperature
factors. Additionally, the relative electron densities of $S_2$ and $S_3$
positions (with respect to $S_1$) were refined. It can be appreciated that the
values found are in a fairly good agreement with those in Tab.~I in
the main text.

\section{Surface oxidation}

Spectroscopic ellipsometry can be employed as an effective tool for
the observation of time dependent changes on the sample surface
which express themselves in surface optical properties that vary over
time. Determination of optical constants relies heavily on the
knowledge of sample's multilayer structure. Our thin layers of CuMnAs
were exposed to air so that the presence of surface oxide layer may be
expected. The following analysis led us to conclude that a few
nanometres thick layer of cuprous oxide forms at a time scale of
days. A freshly grown sample (20~nm thick CuMnAs layer on GaP) was put
into the ellipsometer within two hours after being taken out from the
MBE apparatus. A model structure assuming only the CuMnAs layer and
surface roughness was used to fit experimental data as described in
the main text resulting in the determination of optical parameters of
CuMnAs.

%%%% Table by CF with the values found by XRD using 2D detector samples M061 and M062
%%%% other columns from  P. Wadley et al., J. Appl. Cryst. 46, 1749 (2013)
%%%% lattice parameters of the CuMnAs/GaP film are the average ones determined by DK for a large number of films on GaP 
\begin{table}
  \begin{tabular}{l|cll}
    description:   && CuMnAs/GaAs & CuMnAs/GaP \\ 
                   && Ref. \onlinecite{SI-Wadley:2013_a} & this work \\ \hline
    $a$~[\AA]    && 3.820(10) & 3.853(1) \\
    $c$~[\AA]    && 6.318(10) & 6.276(1) \\
    $z/c$ for $S^2$ && 0.265(1) & 0.259(1) \\
    $z/c$ for $S^3$ && 0.670(3) & 0.664(1) \\ 
    ADP~[\AA$^2$]   && - & 0.028(3) \\ \hline
    $R_B$          && - & 3.97\% 
  \end{tabular}
  \caption{Structural parameters of CuMnAs thin films grown on GaP(001) and
    GaAs(001) substrates. The latter  were converted from values published in
  Ref.~\onlinecite{SI-Wadley:2013_a}.}
  \label{SI-tab-01}
\end{table}

%\begin{table}
%  \begin{tabular}{l|cll}
%    substrate:   && GaP      & GaAs\\ \hline
%    $a$~[\AA] && 3.848(1) & 3.820(1) \\
%    $c$~[\AA] && 6.282(1) & 6.318(1) \\
%    $u$        & & 0.3378   & 0.3300   \\   % Mn -1/2
%    $v$         && 0.2375   & 0.2347
%  \end{tabular}
%  \caption{Structural parameters of the CuMnAs layers grown on GaP
%    substrate. Second column summarises previously
%    published\cite{Wadley:2013_a} data pertaining to the sample used
%    for UPS. Parameters $u,v$ are the $z/c$ values for $S^2$ and $S^3$
%  sites, see Fig.~\ref{fig-04}.}
%  \label{tab-01}
%\end{table}

Afterwards, the sample was exposed to air for about two weeks and the
ellipsometric measurements were repeated several times during this
time interval. The whole set of experimental data (i.e., for four different
times of air exposure) was fitted at once with a modified model structure. This
structure assumed another layer on top of CuMnAs. Only three thicknesses
($l_{\mathrm{CuMnAs}}$, thickness of the additional layer $l_o$ and surface
roughness $l_r$) were left as fitting parameters. The best fit was
obtained using optical parameters of Cu$_2$O and resulting thickness of
this layer ($l_o$) was an increasing function of time elapsed from the
growth. Manganese and arsenic oxides yielded clearly worse fits. Since
the surface roughness in the fit was treated as an effective layer
containing 50\% of air and 50\% of the oxide, the total effective thickness
was calculated as $l_{\mathrm{CuMnAs}}+l_o+\frac12 l_r$ and this quantity
becomes  larger in a matter of days, as shown in Fig.~\ref{SI-fig-01}, and
agrees very well (perhaps even surprisingly well) with the results
of x-ray characterisation.

%Ellipsometric determination of optical constants relies heavily on the
%knowledge of sample's multilayer structure. Our thin layers of CuMnAs
%were exposed to air so that the presence of surface oxide layer may be
%expected. The following analysis led us to conclude that a few
%nanometre thick layer of copper oxide forms at a time scale of days.

\begin{figure}
\includegraphics[scale=0.3]{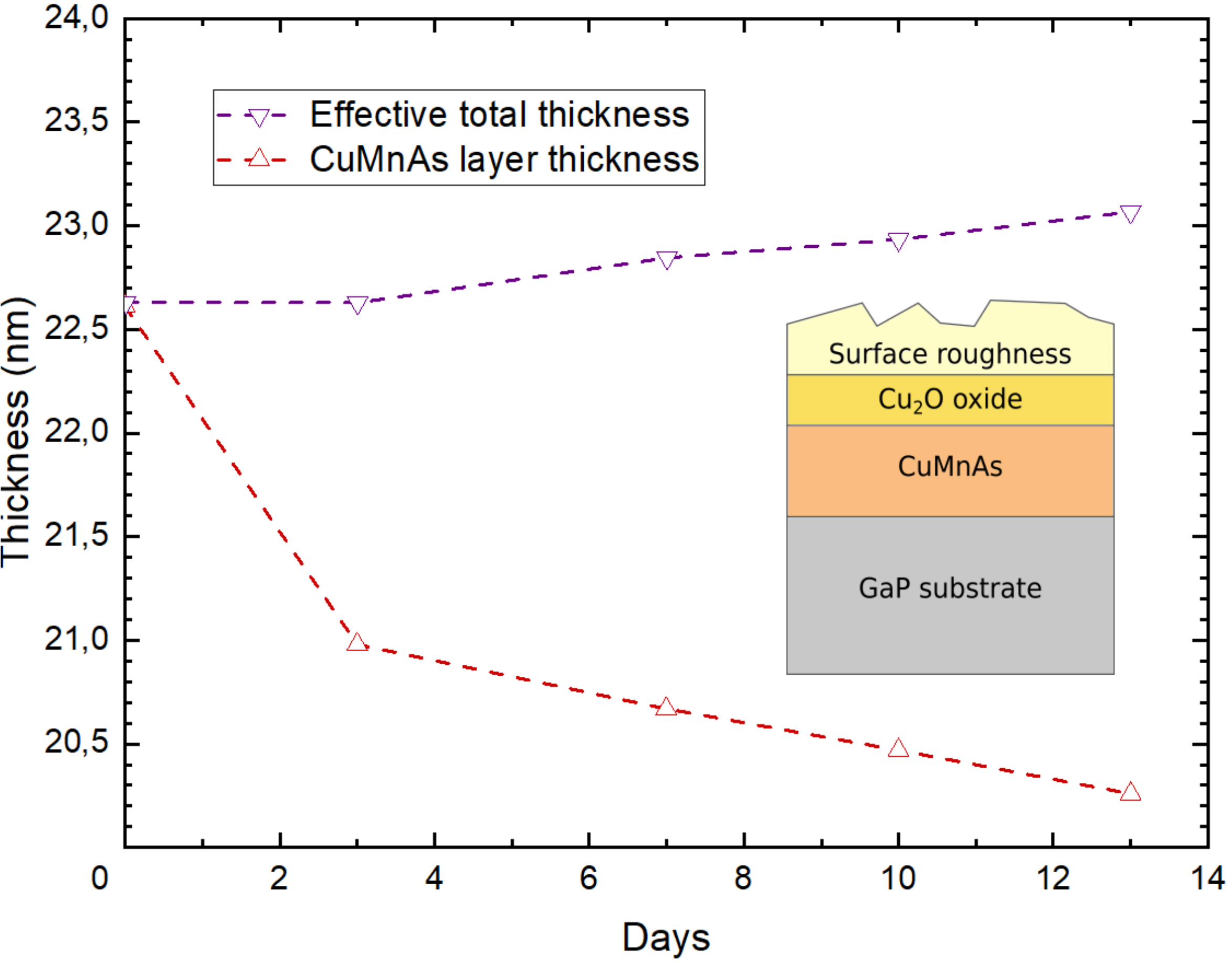}
\caption{After several days of air exposure (starting after the sample
  is taken out from MBE chamber), a thin oxide layer forms on the surface,
  gnawing at the actual material (CuMnAs). Nevertheless, the effective
  total thickness of the layer (see definition in the text) still slightly
  increases with time.}
\label{SI-fig-01}
\end{figure}

%A freshly grown sample (20~nm thick CuMnAs layer on GaP) was put into
%the ellipsometer within hours after being taken out from the MBE
%apparatus. Fitting assuming only the CuMnAs layer and surface
%roughness was performed as described in the main text and optical
%parameters of CuMnAs were obtained. Next, this procedure was repeated
%in an interval of several days with some modification: on top of CuMnAs,
%another layer was assumed to be present and all three thicknesses
%($l_{\mathrm{CuMnAs}}$, thickness of the additional layer
%and surface roughness $l_r$) were left as fitting
%parameters. Good fits were obtained using optical parameters of
%Cu$_2$O and resulting thickness of this layer ($l_o$) was an increasing
%function of time elapsed from the growth. Manganese and
%arsenic oxides yielded clearly worse fits. In Fig.~\ref{fig-01}, we
%show that even though such obtained $l_{\mathrm{CuMnAs}}$ decreases, 
%the total effective thickness calculated as
%$l_{\mathrm{CuMnAs}}+l_o+\frac12 l_r$, becomes larger in a matter of
%days and agrees with the results of x-ray characterisation.

\section{Orthorhombic sample}

The imaginary part of permittivity shown by dashed line in Fig.~2 of
the main text was measured on a bulk sample of CuMnAs grown
using Bi flux similar to previously reported
procedure\cite{Emmanouilidou:2017_a,Uhlirova:2015_a} for the growth
of CuMn$_3$As$_2$ and Cu$_2$Mn$_4$As$_3$.
The elements with starting molar ratio Cu:Mn:As:Bi was 1:1:1:10
(1:1:1:15 also gives good results) were put into alumina crucibles
(diameter 12~mm) and sealed under vacuum ($\sim 10^{-6}$~mbar)
in quartz glass ampoules. The samples were slowly (2~$^\circ$/min)
heated up to 850~$^\circ$C and kept at fixed temperature for 10~h.
Then, they were slowly cooled down to 400~$^\circ$C (3~$^\circ$C/h)
where the Bi flux was centrifuged. The single crystals were
typically 1~to~3~mm long and 50~to~300~$\mu$m thick
with the mass under 0.5 mg. The composition of the prepared single
crystals was determined using energy dispersive x-ray (EDX)
analysis.

Single crystals with composition Cu$_{34}$Mn$_{33}$As$_{33}$
were then subject of further studies. The variation in composition in
selected samples was under 0.5\%. Ellipsometry was performed by the
same procedure as for tetragonal films. The crystal symmetry and lattice
parameters were determined by single crystal x-ray diffraction (XRD)
using Rigaku RAPID II with Mo-K$\alpha$ radiation in transmission
geometry. We have confirmed orthogonal structure (space group Pnma)
with lattice parameters $a=0.6598(4)$~nm, $b=0.3861(4)$~nm and
$c=0.73015(1)$~nm. The lattice parameters and magnetic behaviour of
prepared samples are in agreement with recently reported results on
single crystals prepared by the same
method.\cite{Emmanouilidou:2017_a} 

Typical orthorhombic CuMnAs single crystal is shown in Fig.~\ref{SI-fig-05}. The
remaining Bi flux does not wet the surface and forms small
droplets. For the ellipsometry measurements presented in this work,
rather than etching and exposing the surface to water, we used fresh
as grown surface and selected an area free of the bismuth droplets
(focused beam with spot size $0.15$ to $0.2$~mm was used).

\begin{figure}
\includegraphics[width=0.7\linewidth]{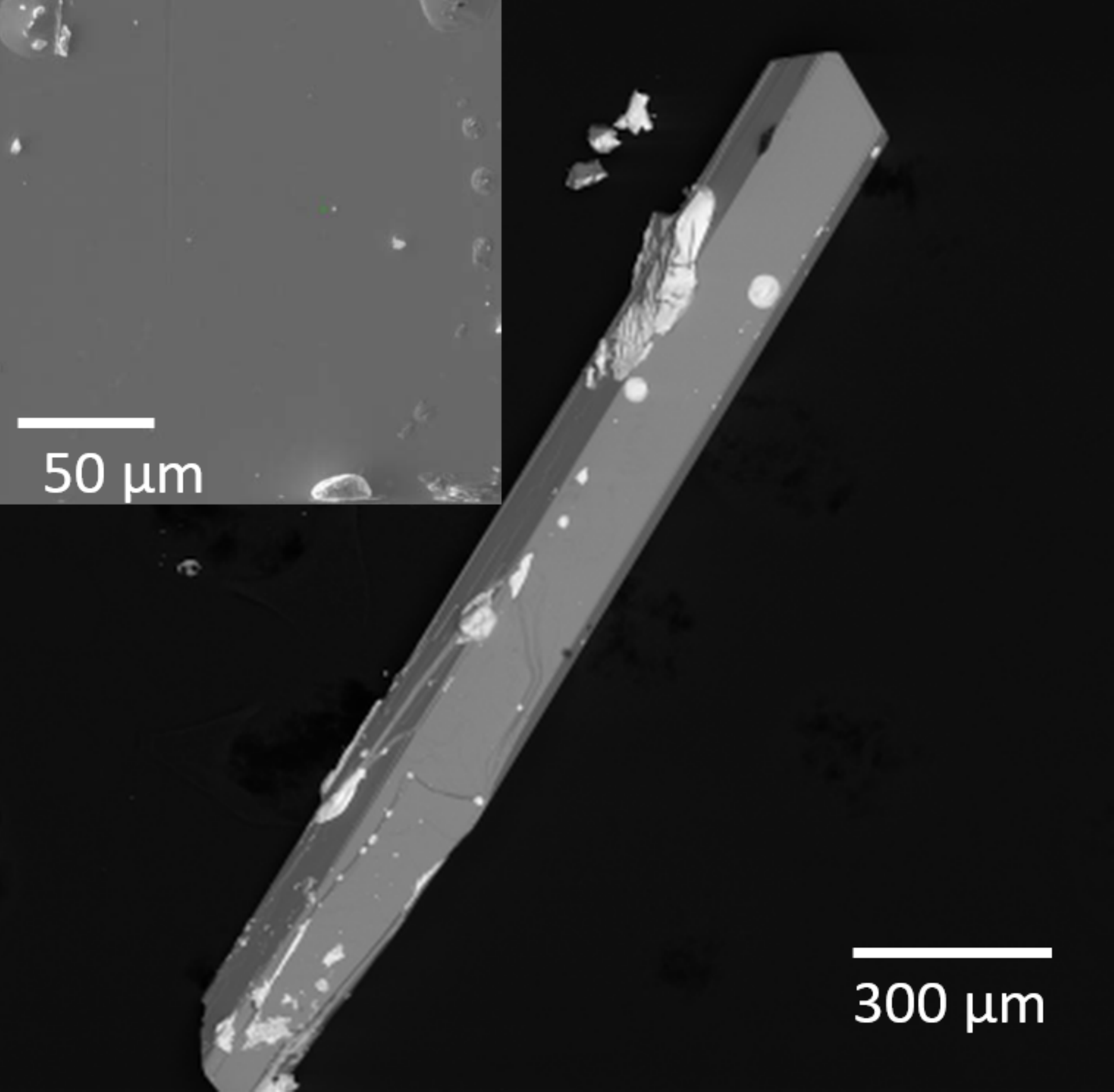}
\caption{Image of orthorhombic CuMnAs single crystals taken by SEM
microscope. In BSE contrast, the remaining Bi flux droplets are bright
and clearly visible. The inset shows detail (SE contrast) on the
sample surface with a clean area suitable for ellipsometric
measurements.}
\label{SI-fig-05}
\end{figure}

\section{UPS modelling and IPES}

In oder to describe photoemission spectra, we used the so called
one-step model of photoemission.  We used recently developed
fully spin-density matrix formulation for the photocurrent\cite{BMK+14}
within the relativistic Korringa–Kohn–Rostoker (SPR-KKR)
Green function method.  As a first step of our PES investigations,
we performed self-consistent LSDA+U ground state calculations for
CuMnAs by means of SPR-KKR method. All parameters of the calcualtions
has been as far as possible same as for the above 
mentioned LAPW  based investigations and results obtained within
SPR-KKR method are quantitativaly comparable to the LAPW method. 
The self consistent potentials and LSDA+U self energy are then used as
an input for UPS and IPE investigations of CuMnAs(001) surface. 
 As the LSDA+U does not include many-body finite life-time of the initial
state this effect was included phenomenologically by imaginary part of
potential (0.05~eV). The impurity scattering of the final state and its
innelastic mean free path was modelled again by the imaginary part of
the inner potential (2.0~eV) as usual.\cite{Pen74}

%... the LSDA+U does not include many-body finite life-time of the initial
%state, this effect was included phenomenologically by a nonzero
%imaginary part of the potential (0.05~eV). The impurity scattering of
%the final state and its inelastic mean free path was again
%modelled\cite{Pen74} by imaginary part of the inner potential. All
%calculated spectra have been broadened by gaussian function
%(FWHM=0.8~eV) in order to simulate experimental resolution.
%Several models of the surface barrier were considered and feature (B)
%of the spectra in Fig.~5 of the main text was thus found to be
%surface-related (see the left panel of Fig.~\ref{fig-06}).

\begin{figure}
  \begin{tabular}{cc}
  \includegraphics[width=0.5\linewidth]{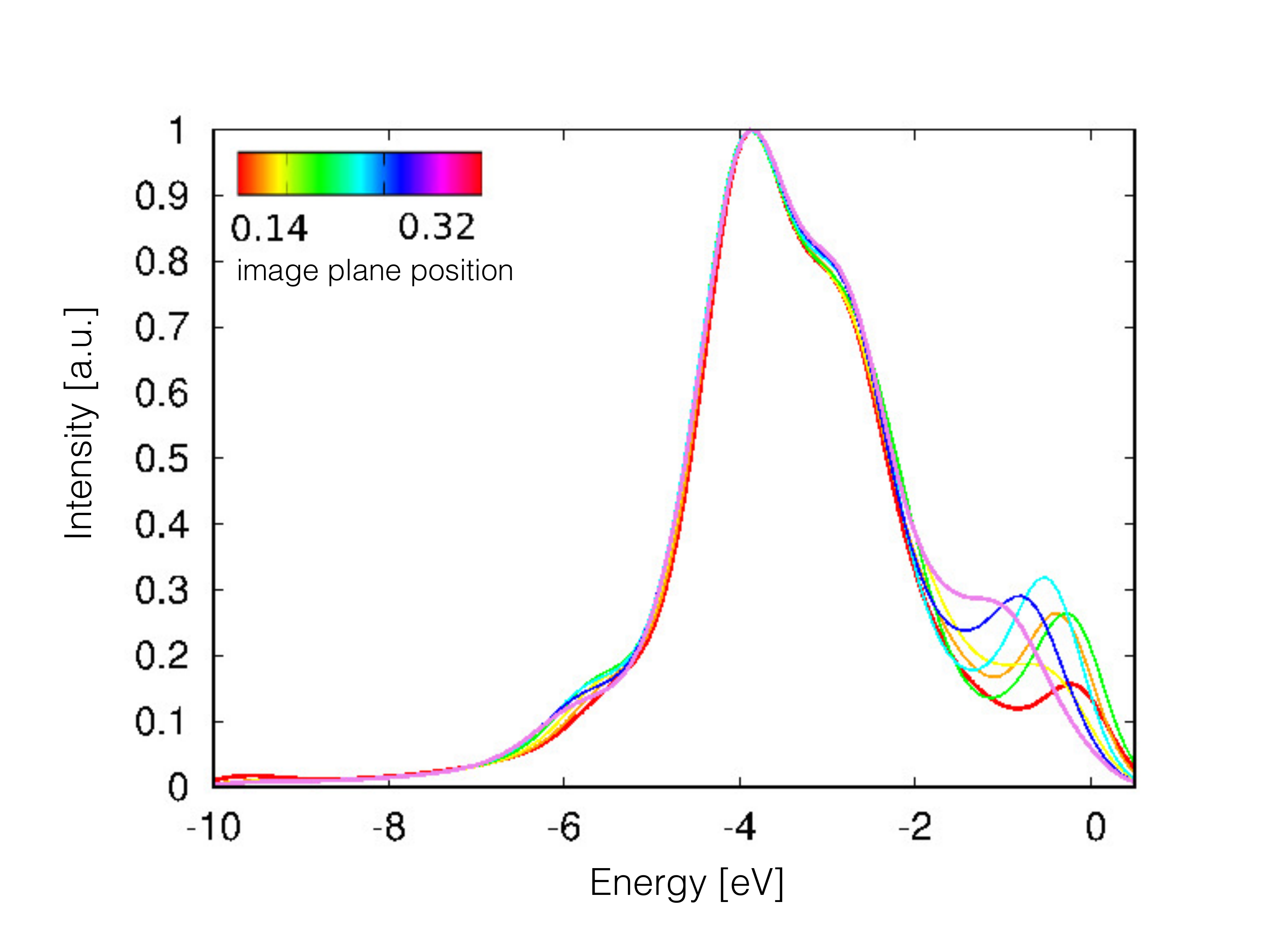} &
  \includegraphics[width=0.53\linewidth]{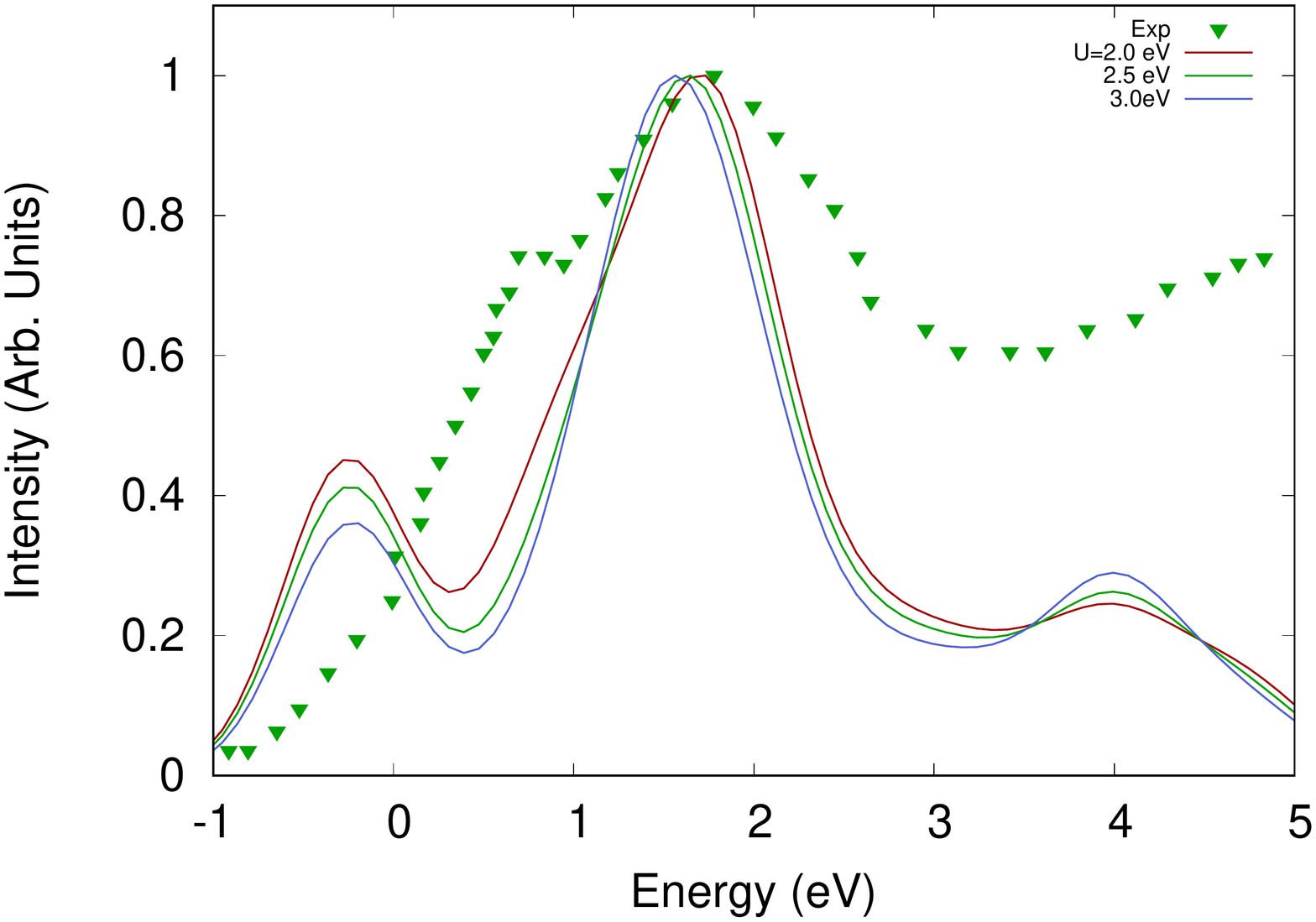}
  \end{tabular}
  \caption{{\em Left:} Result of one-step model of UPS ($U=3$~eV) for
    different models of the surface 
    barrier. {\em Right:} IPES measurement and modelling (RTP).} 
\label{SI-fig-06}
\end{figure}

Additionally, for the photoemission calculations, we accounted for the
surface barrier by use of a Rundgren-Malmstr\"om surface potential
\cite{MR80}, which is included as an additional layer. This procedure
is described in  for example in Ref.~\onlinecite{NBF+11} and
it accounts for the energetics
and dispersion of all surface features. One of the most important
parameters for this surface potential is the position of the classical
image plane which in other words describes distance between surface
barrier  and last surface layer. In the left panel of Fig.~\ref{SI-fig-06},
we show UPS spectra calculated for the position of the image plane
in the range between $0.14c$ and $0.32c$. % $0.14a_{lat}--0.32a_{lat}$.  
The spectral feature close to the Fermi level (B in Fig.~5 of the main
text) clearly shows a strong surface character. On the other hand, the
main peak at binding energy of $-4$~eV (feature A of Fig.~5)
are bulk states with predominant Mn-character.

The conduction band electronic structure of the interfaces was studied
by means of Inverse PhotoEmission Spectroscopy (IPES). Spectrum after
integral background subtraction\cite{SI-Cantoni:1999_a} is
shown in the right panel of Fig.~\ref{SI-fig-06}. The apparatus
consists of an electron source based on a negative electron affinity GaAs
photocathode coupled to an appropriate transport electron optics and a
bandpass out-coming photons detector at fixed energy (9.3~eV)
employing a KBr photocathode and a SrF$_2$ window.
%{\em It shows three main peaks C, D and E located at
%0.85~eV, 1.85 eV and 5.1 eV  above the Fermi energy,
%respectively.}
Like in the case of UPS, the Gaussian instrumental resolution
broadening has been taken into account with $\mbox{FWHM}=0.8$~eV (a
conservative estimate\cite{SI-Cantoni:1999_a}),
evaluated measuring the IPES Fermi-edge of a monocrystalline Ag
sample. In this case, the instrumental broadening is so high that it
has not been necessary to consider the energy spreading due to
lifetime effects. Agreement between experimental data and
one-step model of photoemission is good, again there is one dominant
peak (close to energy of 2~eV) which matches well the calculations
whose dependence on Hubbard $U$ is weaker than in the case of
UPS. Spectral features below the energy of 1~eV are related to surface states.

%The result is shown in Fig. 1(b): the DOS calculated
%for U=0.31~Ry (blue dashed line) resemble the shape of the
%experimental spectrum. Once the instrumental broadening is considered
%(red solid line), both for the intensity and the position of peaks C
%and D are well fitted. For the peak E the agreement is less
%pronounced, probably due to surface effects and lifetime broadening. 

%\section{Analysis of the inverted structure}
%
%UPS/IPES in Fig.~\ref{fig-03} suggests that $U\approx 3$~eV. The peak
%in Im~$\epsilon(\omega)/\epsilon_0$ is then about half an eV too low.
%
%\begin{figure}
%\includegraphics[width=0.7\linewidth]{figs/p03.pdf}
%%\includegraphics[width=0.7\linewidth]{p03.pdf}
%\caption{UPS/IPES for the inverted structure.} 
%\label{fig-03}
%\end{figure}

\end{widetext}

\end{document}